\documentclass[aps,twocolumn,superscriptaddress,showpacs,pre,floatfix]{revtex4-1}
\usepackage{graphicx,color}

\def\costanteFlusso{F}
\def\frazione{\xi}

\begin{document}

\title{Which is the flavor of cosmic neutrinos seen by IceCube?}

\author{A. Palladino}
\affiliation{Gran Sasso Science Institute, L'Aquila (AQ), Italy}
\author{G. Pagliaroli}
\affiliation{INFN, Laboratori Nazionali del Gran Sasso, Assergi (AQ), Italy}
\author{F.L. Villante}
\affiliation{L'Aquila University, Physics and Chemistry Department, L'Aquila, Italy}
\affiliation{INFN, Laboratori Nazionali del Gran Sasso, Assergi (AQ),  Italy}
\author{F. Vissani}
\affiliation{INFN, Laboratori Nazionali del Gran Sasso, Assergi (AQ), Italy}
\affiliation{Gran Sasso Science Institute, L'Aquila (AQ), Italy}

\begin{abstract}

We analyze the high-energy neutrino
events observed by IceCube, aiming to probe 
the initial flavor of cosmic neutrinos.
We study the track-to-shower ratio of the subset 
with energy above 60 TeV, where the signal is expected to dominate
and show that different production mechanisms 
give rise to different predictions even accounting for the uncertainties due to neutrino oscillations. 
We include for the first time the passing muons observed by IceCube in the analysis. They corroborate the hypotheses that cosmic neutrinos have been seen and their flavor matches expectations. 
\end{abstract}
\pacs{95.85.Ry, 14.60.Pq, 95.55.Vj, 29.40.Ka}
\maketitle

\section{Introduction}

The search for High Energy Starting Events (HESE) in IceCube detector
provided the first evidence for a high-energy neutrino flux 
of extraterrestrial origin
\cite{IceCube3years,IceCube1TeV,IceCubeScience}. 
In three year of data taking \cite{IceCube3years}, 
37 events with deposited energies above 30 TeV 
were observed, relative to an expected
background of $8.4 \pm 4.2$ cosmic ray muon events and
$6.6\pm $5.9 atmospheric neutrinos. 

The scientific debate about the origin of these events  
is extremely lively.
There is little doubt that cosmic neutrinos have been seen, 
but their origin and propagation is not understood.
In order to proceed, the flavor composition has to be probed.
The flavor discrimination is, in principle, possible by looking at 
the topology of the events.
Most HESE  have `shower' topology, that includes  neutral current (NC) 
interactions of all neutrino flavors and
charged current (CC) interactions of 
$\nu_{\mathrm{e}}$  and $\nu_\tau$, being the decay length of the $\tau$ lepton 
too short to be resolved below $\sim 1 {\rm PeV}$. 
%
On the other hand,
events with `track' topology are produced by CC interactions of $\nu_\mu$.
Thus, the crucial observable quantity is the ratio between 
track and shower events at high energy and it can be used 
to confirm the cosmic origin and/or to discriminate 
among different production scenarios.
With this purpose the authors of \cite{Mena,Mena2} recently 
discussed the observed track-to-shower ratio of the IceCube data 
with energy above 30 TeV  claiming that these data are 
marginally compatible with the hypothesis that neutrinos 
are of cosmic origin. These studies have been
influential, setting the case for a muon deficit problem in IceCube, 
see e.g.\ \cite{Anchordoqui2014,Chen2014}.

In view of the importance of this issue, we perform an
independent analysis adding our contribution to the discussion. 
We focus  on the subset of events with deposited 
energy  above 60 TeV, where the signal is expected to dominate. 
We show that different production mechanisms give rise to distinctive 
expectations of the track-to-shower ratio, 
even when the uncertainties due to neutrino oscillations are included. 
Also the  muon neutrinos
passing through the Earth 
confirm the existence of an
astrophysical component and we include
for the first time 
 this information  on the analysis. 
We find that present data set 
is well compatible with the hypothesis that cosmic 
neutrinos have been seen, even if  the limited statistics does not allow yet to discriminate the initial flavor.

\section{Expectations}
\paragraph*{From neutrinos to HESE events.}
Let us consider HESE events with energies between 60 TeV to 3 PeV and
starting inside IceCube, that are likely to be dominated by the signal due to cosmic neutrinos. 
The expected number of events produced by an isotropic flux $\Phi_{\ell}$ 
of neutrinos and antineutrinos with flavor $\ell$ is,
\begin{equation}
N  = 4\pi \, T \int dE \; \Phi_{\ell} (E) \,  A_{\ell}(E)
\end{equation}
where $\ell = e,\,\mu,\,\tau$ and $T$ is the observation time.
The effective areas $A_{\ell}(E)$  are provided by the IceCube 
collaboration \cite{IceCubeScience} and include the effects of neutrino
cross sections, partial neutrino absorption in the earth, detector efficiency 
and specific cuts of the HESE analysis.

In order to calculate the track-to-shower ratio,
we separate the different contributions 
to the effective areas,
\begin{equation}
A_{\mu}(E) = A^{\rm T}_{\mu}(E) + A^{\rm S}_{\mu}(E)
\end{equation}
where $A^{\rm T}_{\mu} (E) \equiv p_{\rm T}(E) \, A_\mu (E)$ is the
effective area for $\nu_\mu$ CC interactions that produce tracks in the detector, 
while $A^{\rm S}_{\mu}(E) \equiv (1-p_{\rm T}(E))\, A_\mu (E)$ is the
effective area for neutral current (NC) interactions that are instead observed as showers. 
The parameter $p_{\rm T}(E)$ is the probability
that an {\em observed} event 
(i.e.\ passing all the cuts in the HESE analysis) produced by a muon
neutrino with energy $E$ is a track event.
This quantity is given by,
$$
p_{\rm T}(E) = \frac{\sigma_{\rm CC}(E) \, M^{\rm CC}_{\mu}(E) }
{\sigma_{\rm NC}(E) \, M^{\rm NC}(E) + \sigma_{\rm CC}(E) \,
  M^{\rm CC}_{\mu}(E)} 
$$
where $\sigma_{\rm CC}(E)$ and $\sigma_{\rm NC}(E)$ are the cross
section for CC and NC interactions of neutrinos~\cite{crosssections}
while  $M^{\rm CC}_{\mu}(E)$  and $M^{\rm NC}(E)$ are the
effective detector mass for CC and NC interactions 
of $\nu_\mu$ \cite{IceCubeScience}. The probability $p_{\rm T}$ 
is mildly dependent on energy and approximately equals 
$0.8$. 

The number of showers $N_{\rm S}$ and tracks $N_{\rm T}$ in the
IceCube detector can be then calculated according to:
\begin{eqnarray}
\nonumber
N_{\rm S} &=& 4\pi \, T \int^{\overline{E}}_{0} dE \; \left\{ 
\Phi_{e} (E) \,  A_{e}(E) + \Phi_{\tau} (E) \,  A_{\tau}(E)  + \right.\\
\nonumber
&+& \left. \Phi_{\mu} (E) \, \left[1- p_{\rm T}\right] \,  A_{\mu}(E) \right\} \\
N_{\rm T} &=& 4\pi \, T \int^{\overline{E}}_{0} dE \; 
 \Phi_{\mu} (E)  \,  p_{\rm T}  \, A_{\mu}(E) \,  
\label{NTandNS}
\end{eqnarray}
In the above relation, we neglected the small fraction of $\nu_\tau$ CC-events
followed by taus decaying into muons which can be potentially observed as tracks.
Moreover, we introduced an upper integration limit at $\overline{E} =
3\,{\rm PeV}$
since the HESE analysis only includes events with deposited energy
below $3 \,{\rm PeV}$. 
In principle, the effects of the threshold at $E_{\rm dep} = 3\,{\rm PeV}$ 
should be implemented as a correction of the effective areas. 
Here, we assume that this can be mimicked by a sharp cut in the
$A_{\ell}(E)$ at the neutrino energy $E = 3\,{\rm PeV}$. 
We tested the validity of this approach by comparing our
predictions 
with the expected numbers of events 
in Supp. Tab. IV of \cite{IceCube3years}. We obtain good agreement
both for the absolute and relative numbers of shower and track events.

\paragraph*{Description of cosmic neutrinos.}
Cosmic neutrinos are surely due to non-thermal processes. Thus we expect that their fluxes 
averaged  over the directions, are approximated by a power law
distributions up to a maximum value that we
assume being larger than 3 PeV, 
\begin{equation}
\Phi_\ell(E) =\frac{\costanteFlusso_\ell \cdot
  10^{-8}}{\mbox{cm$^2$ s sr GeV}} \cdot
\left(\frac{\mbox{GeV}}{E}\right)^\alpha 
\label{fluxes}
\end{equation} 
where 
the factors $\costanteFlusso_\ell$ are (non-negative) adimensional coefficients 
and  $\alpha$ is the spectral index.
We use the value $\alpha =2$, expected on theoretical basis, 
and find the following 
expressions for number of  shower and track events,
\begin{eqnarray} 
\nonumber
N_{\rm S} &=& 8.4\times  \costanteFlusso_e + 0.9 \times  \costanteFlusso_\mu+6.3\times  \costanteFlusso_\tau\\
N_{\rm T} &=& 3.7\times  \costanteFlusso_\mu 
\label{NTandNsSimple}
\end{eqnarray}
The track-to-shower ratio is  then,
\begin{equation}
\frac{N_{\rm T}}{N_{\rm S}} = \frac{\frazione_\mu}{2.3 -  2.0 \, \frazione_\mu -
  0.6 \, \frazione_\tau}
\label{NToverNs}
\end{equation}
where we introduced the {\em flavor fractions at  Earth} (i.e., in the detection point), 
defined as:
\begin{equation}\label{ciups}
\frazione_\ell\equiv \costanteFlusso_\ell/\costanteFlusso_{\rm tot}
\end{equation}
with $\costanteFlusso_{\rm tot} = \costanteFlusso_e+\costanteFlusso_\mu+\costanteFlusso_\tau$,
and we considered that  $\frazione_e = 1 - \frazione_\mu -
\frazione_\tau$.
The numerical coefficients of  eq.~(\ref{NToverNs}) 
 depend mildly on the spectral index, as  
quantified later.


\paragraph*{The effect of neutrino oscillations.}
For neutrinos travelling over cosmic distances, the simplest regime
(Gribov-Pontecorvo's \cite{gp}) 
applies and the oscillation probabilities $P_{\ell\ell'}$ are energy
independent.  The flavor fractions at Earth are thus given by  
$$
\frazione_\ell 
= \sum_{\ell'} P_{\ell\ell'} \, \frazione_{\ell'}^0 
\;\;\;\;\;
\mbox{with} 
\;\;\;\;\;
P_{\ell\ell'}=\sum_{i=1,3} |U_{\ell i}^2 \ U_{\ell' i}^2 |,
$$
where $U$ is the neutrino mixing matrix and 
$\frazione_{\ell}^0$ are the {\em flavor fractions at the source}
given by:
\begin{equation} 
\frazione^0_\ell\equiv \costanteFlusso^0_\ell/\costanteFlusso_{\rm tot}
\end{equation}
where $\costanteFlusso_{\ell}^0$ indicates the adimensional flux 
normalizations before oscillations--see eqs.~(\ref{fluxes},\ref{ciups}). 
It is generally expected, see e.g.\ \cite{athar,Noi,Others} that  
a cosmic population is characterized 
by a flavor content $(\xi_e:\xi_{\mu}:\xi_{\tau}) \sim (1/3 : 1/3 : 1/3)$
independently on the specific production mechanism. 
In this case, the track to-shower ratio in IceCube is,
%
\begin{equation}
\frac{N_{\rm T}}{N_{\rm S}} = 0.24
\label{equi}
\end{equation}
as can be calculated from eq.~(\ref{NTandNsSimple}). If we consider a
spectral index $\alpha\neq 2$, this prediction is only marginally affected being 
approximately $N_{\rm T}/N_{\rm S} = 0.24+0.08\,(2-\alpha)$.
%
%

The equipartition of neutrino flavors at Earth 
is, however, only an approximation 
which is no longer adequate after IceCube data:
A certain imprint of the neutrino production mechanism 
does remain. It is important to exploit  
the track-to-shower ratio observed by IceCube 
to discriminate neutrino origin.
To explore this possibility on realistic grounds, it is necessary to quantify
the relevance of uncertainties in oscillation parameters
for the predictions of $N_{\rm T}/N_{\rm S}$. 
We note that the probabilities $P_{\ell\ell'}$ have a non-linear dependence 
on the neutrino oscillation parameters 
and, as a consequence, the errors in $\theta_{12} $, $\theta_{13} $,
$\theta_{23}$ and $\delta$ cannot be propagated linearly. 
Moreover, the allowed regions for $\theta_{23}$ and $\delta$ parameters have 
complicated structures that cannot be correctly described by assuming gaussian dispersions 
with the quoted $1\sigma$ errors.
We overcame these difficulties by constructing likelihood distributions of
$\sin^2 \theta_{12} $, $\sin^2 \theta_{13} $, $\sin^2 \theta_{23}$ and $\delta$
from the $\Delta\chi^2$ profiles given by \cite{nufit}. Namely, we assume
that the probability distributions of each parameter are provided by
${\mathcal L} = \exp{\left(-\Delta \chi^2/2\right)}$. Then, we combine
the various likelihood functions assuming negligible correlations
and we determine the probability distributions of $N_{\rm T}/N_{\rm S}$ 
by MonteCarlo extraction of the oscillation parameters.
We consider 
four specific assumptions for the flavor composition at the source
$(\frazione^0_e :\frazione^0_{\mu} :\frazione^0_{\tau})$ 
 which are relevant for the interpretation 
of observational data because related to specific production
mechanisms. We consider\\
{\em i)}  $(1/3 : 2/3 : 0)$ for $\pi$ decay (yellow);\\
{\em ii)} $(1/2 : 1/2 : 0)$ for  {\em charmed mesons} decay (blue); \\
{\em iii)} $(1 : 0 : 0)$ for $\beta$ decay of {\em neutrons} (green);\\
{\em iv)} $(0 : 1 : 0)$ for $\pi$ decay with {\em damped
  muons} (red),\\
  where we made reference to the color code
used in Fig.~\ref{fig1}.

Fig.~\ref{fig1} summarizes our results. The left panel is obtained by using the distribution of the 
oscillation parameters corresponding to the assumption of normal hierarchy (NH), while
the right panel corresponds to the case of inverse hierarchy (IH). 
We see that $N_{\rm T}/N_{\rm S}$ distributions are well separated when 
different neutrino production mechanisms are considered. This 
means that a precise determination of
$N_{\rm T}/N_{\rm S}$ could provide hints on the neutrino origin,
even with the present knowledge of neutrino mixing parameters. 
From the neutrino physics point of view, large contributions to
$N_{\rm T}/N_{\rm S}$ dispersions are provided by the $\delta$ and $\theta_{23}$ parameters. 
Finally, our results indicates that 
the flavor composition of cosmic neutrinos cannot be used to learn about neutrinos, 
unless the neutrino production mechanism is independently identified.

For the purposes of our discussion, it is finally important to note 
that the track-to-shower ratio has a limited
range of possible values, if neutrinos have cosmic origin. 
If we take the best fit oscillation parameters and assume a
spectral index $\alpha =2$, we obtain 
\begin{equation}
0.15 < \frac{N_{\rm T}}{N_{\rm S}} < 0.30 \mbox{\ \ \ \ [expected from cosmic origin]}
\label{2}
\end{equation}
The minimal value, obtained for 
neutron-decay (i.e.\ $\frazione^0_\mu=\frazione^0_\tau=0$ and $\frazione^0_e =1$), 
matters for the claims of 
a possible muon deficit problem in IceCube. 
If we vary the spectral index, this interval shifts by $\sim \mp$~10\%. 
The oscillation parameters affect slightly these expectations; 
e.g.\ for the lowest (resp.\ highest)  value of
$\sin^2\!\theta_{23}=0.385$ (resp.\ =0.644)~\cite{nufit}, 
the interval of Eq.~(\ref{2}) narrows to [0.16,0.27] (resp., widens to [0.09,0.43]).

\section{Data analysis}

\paragraph*{General considerations.}

In the energy window $60 \,{\rm TeV} < E_{\rm dep} < 3 \,{\rm PeV}$, 
20 events have been observed, consisting of 16 showers 
and 4 tracks events, against an expected background of $\sim 3$ events 
from atmospheric muons and neutrinos.
By performing a likelihood fit, an isotropic astrophysical component 
with $E^{-2}$ spectrum and flavor composition
$(1/3: 1/3: 1/3)$, as expected due to neutrino flavor oscillations
(see e.g.\ \cite{athar,Noi,Others}), 
is extracted at $5.7\, \sigma$ confidence level \cite{IceCube3years}. 
Namely, the best fit astrophysical neutrino flux is given by
$E^2 \, \Phi_\ell(E) = ( 0.95 \pm 0.3 ) \times 10^{-8} 
\, {\rm GeV} \, {\rm  cm}^{-2} \, {\rm s}^{-1} \, {\rm sr}^{-1}$, where the index 
$\ell= e,\,\mu,\,\tau$ refers to the neutrino flavor.

 New data and analyses confirm the evidence for 
a cosmic neutrino population. Recently, a new technique was 
developed that permits to isolate events starting in the IceCube 
detector down to $\sim 1 {\rm TeV}$ and 
to observe astrophysical neutrinos (in the southern sky) with energies
as low as 10 TeV \cite{IceCube1TeV}. 
Even more interesting, an independent analysis 
of the spectrum of muon neutrinos passing through the Earth
has confirmed the existence of an astrophysical component. 
Analyzing the same period of the HESE analysis, an excess of
high energy muon tracks is observed, that was fitted 
by assuming an astrophysical muon neutrino flux equal to
$E^2 \, \Phi_\mu(E) = ( 1.01\pm 0.35 ) \times 10^{-8} 
\, {\rm GeV} \, {\rm  cm}^{-2} \, {\rm s}^{-1} \, {\rm sr}^{-1}$
\cite{IceCubePassingMuons, IceCubeVision}.

\begin{figure*}[th]
\par
\begin{center}
\includegraphics[width=.9\textwidth,angle=0]{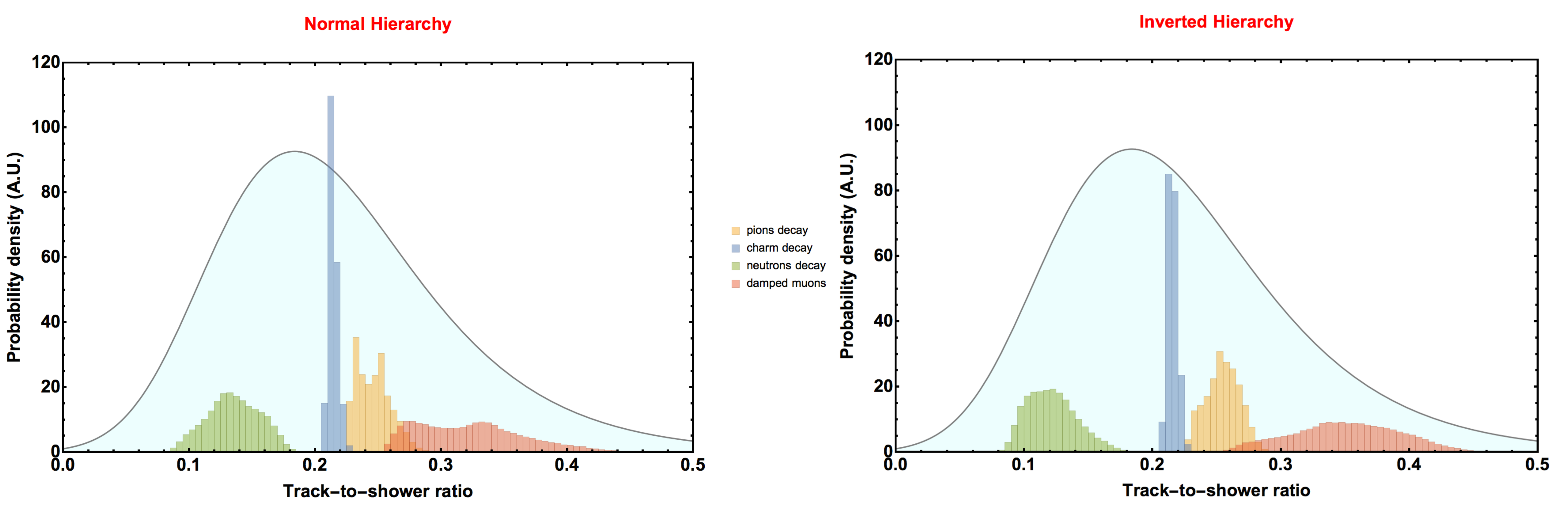}
\end{center}
\par
\vspace{-7mm} \caption{\em\protect\small  
Expected track-to-shower ratio of cosmic neutrinos
for the four production mechanisms described in the text. The distributions show the 
effect of uncertainties in the neutrino  oscillation parameters. 
The left (resp.\ right) panel is obtained for normal (resp.\ inverse) hierarchy. 
The shaded  region is the likelihood corresponding to Eq.~(\ref{1}).}
\label{fig1}
\end{figure*}

\paragraph*{The track-to-shower ratio.}
The set of events observed by IceCube in three years of data taking 
between 60 TeV and 3 PeV consists of  a total number of $n_{\rm T} =4$
tracks and $n_{\rm S} = 16$ showers.  These includes on average $b_{\rm T} = 2.1 $ and  $b_{\rm S} = 0.7$ 
background events expected from atmospheric neutrinos (1.7 tracks and 0.7 showers) and muons 
(0.4 tracks and no showers) \cite{IceCube3years}. 
In the above estimates, we assume that the prompt atmospheric 
neutrinos give negligible contributions, as it required by the
spectral and arrival angles distributions of IceCube events. 
The number of tracks $N_{\rm T}$ and showers  $N_{\rm S}$ which have to be ascribed to cosmic sources 
can be estimated from the Poisson likelihood functions:
$
\mathcal{L}(N_{\rm i}) \propto
\lambda_{\rm i}^{n_{\rm i}} \times  e^{-\lambda_{\rm i}}
$ 
where $\lambda_{\rm i}=N_{\rm i} + b_{\rm i}$ and the index ${\rm i}={\rm T,S}$ is used to
refer to track and shower events. 
By using the above data, we obtain $N_{\rm T}=3.1\pm 2.1$ and $N_{\rm S}=16.3\pm 4.1$.
Marginalizing over the total number of events, we reconstruct the
track-to-shower ratio of cosmic neutrino obtaining
\begin{equation}
\frac{N_{\rm T}}{N_{\rm S}} = 0.11 ^{+0.23}_{-0.05}
\mbox{\ \ \ \  [HESE only]} 
\label{HESE} 
\end{equation}
where the error was obtained by integrating out symmetrically $(1 -{\rm CL})/2$ on both
sides of the $N_{\rm T}/N_{\rm S}$ distribution using a confidence level 
${\rm CL} =68.3\%$.
The above result can be compared with the range given in eq.~(\ref{2}) 
and shows that IceCube results 
do not contradict the assumption of a cosmic neutrino population. 
The large error in the reconstructed $N_{\rm T}/N_{\rm S}$ is due to
the total number of tracks which is too low to drive any conclusions 
about neutrino origin. Luckily, a completely equivalent and
independent information can be obtained by the recently released
IceCube data on passing muons \cite{IceCubePassingMuons}. 
About $12$ events with visible energy
above $60\,{\rm TeV}$ have been observed which cannot be explained
by atmospheric neutrinos and muons. In the assumption of 
$E^{-2}$ neutrino spectrum, this corresponds to a flux normalization  
 $F_\mu=1.01\pm 0.35$
that can be translated into a number of tracks from cosmic
neutrinos by using eq.~(\ref{NTandNsSimple}). 
We obtain $N_{\rm T}=3.7\pm 1.3$ which is perfectly compatible with the 
value $N_{\rm T}=3.1\pm 2.1$ obtained from the HESE analysis, but is affected
by factor $\sim 2$ smaller error. 
 We include  also this information in our analysis by constructing a combined
 likelihood, given by the product of the 2 Poisson likelihoods for
 $N_{\rm T}$ and $N_{\rm S}$ and of the Gaussian likelihood for $F_\mu$.
We then extract the bound:  
\begin{equation}
\frac{N_{\rm T}}{N_{\rm S}}=0.18^{+0.13}_{-0.05}  \mbox{\ \ \ \ [all data]}
\label{1}
\end{equation}
by taking into account the equivalence between $F_\mu$ and $N_T$
expressed by eq.~(\ref{NTandNsSimple}) and 
marginalizing with respect to the total number of events.
The likelyhood distribution for the track-to-shower
ratio of cosmic neutrinos is shown by the shaded region in Fig.\ref{fig1}.

\section{Discussion and summary}

Fig.\ref{fig1} shows clearly that: {\em i)} there 
is no tension between the present observational results and the 
assumption of a cosmic neutrino population, being the central 
observational value in the middle of the expected region; 
{\em ii)} there is no clear preference for a specific neutrino production 
mechanism, being the observational error comparable to the 
difference between the various predictions.

Our results are substantially different from those obtained by \cite{Mena, Mena2}. 
This is partly due to the inclusion of  the data on passing muons
\cite{IceCubePassingMuons}, 
partly to the fact that \cite{Mena, Mena2} 
include in their analysis the HESE IceCube data between 30 TeV and 60 TeV. 
Following IceCube, we do not consider this region which 
is background dominated and much less valuable to extract the signal.

Below 60 TeV, IceCube observes 16 events, consisting of
4 tracks and 12 showers \cite{IceCube3years}. The sum of tracks and showers
agrees with the expectations but there is a deficit of track events 
(the uncertainty on the background muon rate is, however, 50\%) 
and an excess of shower events. 
If one follows \cite{Mena,Mena2} and supposes that most of the
12 showers are due to cosmic neutrinos, then $N_{\rm S} > 50$ shower events 
are expected above 60 TeV \cite{IceCube3years}, in severe contrast with the 
observational results. In other words, this position is untenable 
if the spectral distribution of the events, not discussed in \cite{Mena}, is considered.
One possible explanation of the track deficit at low energy 
is that few $\nu_\mu$ CC interactions were erroneously identified as
showers since the muon track was missed (e.g., for events occurring close 
to the boundary of the fiducial volume). Our results, 
expressed by Eq.~(\ref{1}), are stable with respect to a possible track
misidentification systematical error. 
Indeed, above 60 TeV, the number of expected showers is much 
larger than the rate of $\nu_\mu$ CC interactions (and thus the
erroneously identified events have a small relative importance on
$N_{\rm S}$). Moreover, $N_{\rm T}$ is well estimated by 
passing muon data \cite{IceCubePassingMuons}
which are free from track misidentification systematics.

To summarize, the HESE events observed by IceCube above 60 TeV are consistent 
with the hypothesis that cosmic neutrinos have been seen.  
The same is true for passing muon events
\cite{IceCubePassingMuons}. The flux of the cosmic
muon neutrinos can be determined reasonably well.   
The analysis of the present data gives a track-to-shower ratio, eq.~(\ref{1}), that agrees with that 
expected for a cosmic population, eq.~(\ref{2}). The initial neutrino flavor cannot 
be yet probed: indeed, all  production mechanisms are allowed.



\begin{thebibliography}{99}



\bibitem{IceCube3years} 
  M.~G.~Aartsen {\it et al.}  [IceCube Collaboration],
  Phys.\ Rev.\ Lett.\  {\bf 113}, 101101 (2014)

\bibitem{IceCube1TeV} 
  M.~G.~Aartsen {\it et al.}  [IceCube Collaboration],
  Phys.\ Rev.\ D {\bf 91}, 022001 (2015)


\bibitem{IceCubeScience} 
  M.~G.~Aartsen {\it et al.}  [IceCube Collaboration],
  Science {\bf 342}, 1242856 (2013)




\bibitem{Mena} 
  O.~Mena, S.~Palomares-Ruiz and A.~C.~Vincent,
  Phys.\ Rev.\ Lett.\  {\bf 113}, no. 9, 091103 (2014)

\bibitem{Mena2} 
  S.~Palomares-Ruiz, O.~Mena and A.~C.~Vincent,
  arXiv:1411.2998 [astro-ph.HE].

\bibitem{Chen2014} 
  C.~Y.~Chen, P.~S.~B.~Dev and A.~Soni,
  arXiv:1411.5658 [hep-ph].

\bibitem{Anchordoqui2014} 
  L.~A.~Anchordoqui,
  arXiv:1411.6457 [astro-ph.HE].

\bibitem{crosssections}
R.~Gandhi, C.~Quigg, M.~H.~Reno and I.~Sarcevic,
  Phys.\ Rev.\ D {\bf 58} (1998) 093009
  
  

\bibitem{gp}
V.~N.~Gribov and B.~Pontecorvo,
  Phys.\ Lett.\ B {\bf 28} (1969) 493. 
  S.~M.~Bilenky and B.~Pontecorvo,
  Phys.\ Rept.\  {\bf 41} (1978) 225.



  

\bibitem{athar}
H.~Athar, M.~Jezabek and O.~Yasuda,
  Phys.\ Rev.\ D {\bf 62} (2000) 103007
  [hep-ph/0005104].
  
  
  



\bibitem{Others}
J.~F.~Beacom,  {\it et al.} 
  Phys.\ Rev.\ D {\bf 68} (2003) 093005
   [E-ibid.\ D {\bf 72} (2005) 019901];
   M.~L.~Costantini and F.~Vissani,
  Astropart.\ Phys.\  {\bf 23} (2005) 477;
  F.~L.~Villante and F.~Vissani,
  Phys.\ Rev.\ D {\bf 78} (2008) 103007.
   
\bibitem{Noi} 
  F.~Vissani, G.~Pagliaroli and F.~L.~Villante,
  JCAP {\bf 1309}, 017 (2013)
  [arXiv:1306.0211 [astro-ph.HE]].





\bibitem{nufit} M.~C.~Gonzalez-Garcia, M.~Maltoni and T.~Schwetz,
  JHEP {\bf 1411}, 052 (2014)


\bibitem{IceCubePassingMuons}
C. Weaver, Spring APS Meeting, Savannah, Georgia (2014).
  M.~G.~Aartsen {\it et al.}  [IceCube Collaboration],

\bibitem{IceCubeVision}
  M.~G.~Aartsen {\it et al.}  [IceCube Collaboration],
  arXiv:1412.5106 [astro-ph.HE].








\end{thebibliography}
\end{document}